\documentclass[preprint,3p]{elsarticle}

\pdfoutput=1
\usepackage{lineno,hyperref}
\modulolinenumbers[5]

\journal{JSV}

\biboptions{numbers,sort&compress} 

\usepackage{mathptmx}       

\usepackage{makeidx}         
\usepackage{graphicx}        
\usepackage{multicol}        
\usepackage{multirow}

\usepackage{enumitem}
\usepackage{natbib}
\usepackage{float}

\usepackage{transparent}

\usepackage{amsmath}
\usepackage{amsfonts}
\usepackage{amssymb}
\usepackage{import}

\usepackage{epsfig}
\usepackage{rotating}
\usepackage{subcaption}
\usepackage{xspace}
\newcommand{\bs}{\boldsymbol}

\newcommand{\define}{:=}

\newcommand{\mbf}{\mathbf}
\newcommand{\real}[1]{\operatorname{Re}\left\lbrace #1 \right\rbrace}

\newcommand{\abs}[1]{\left| #1 \right|}

\newcommand{\ie}{i.e.\,}
\newcommand{\eg}{e.g.\,}
\newcommand{\cf}{c.\,f.\,}
\newcommand{\etal}{et\,al.\,}
\newcommand{\sref}[1]{Section \ref{sec:#1}}
\newcommand{\aref}[1]{\ref{append:#1}}
\newcommand{\eref}[1]{Eq.\ (\ref{eq:#1})}
\newcommand{\fref}[1]{Fig.\ \ref{fig:#1}}
\newcommand{\tref}[1]{Tab.\ \ref{tab:#1}}
\newcommand{\fk}{\,,}

\newcommand{\tra}{{}^{\mathrm T}}
\newcommand{\EPMC}{EPMC\xspace}
\newcommand{\PBMIF}{PBMIF\xspace}
\newcommand{\PBMIFs}{\mathrm{PBMIF}}
\newcommand{\activepower}{P}
\newcommand{\apparentpower}{S}
\renewcommand{\matrix}[2]{\left[\!\!\begin{array}{#1} #2\end{array}\!\!\right]}
\newcommand{\ommod}{\tilde\omega}
\newcommand{\Dmod}{\tilde\delta}
\newcommand{\shpmods}{\tilde{\psi}}
\newcommand{\shpmod}{\tilde{\bs\psi}}
\newcommand{\shpmodnorm}{\tilde{\bs\phi}}
\usepackage{capt-of}
\usepackage{bm}
\usepackage{epstopdf}

\newcommand{\ee}{\mathrm{e}}
\newcommand{\ii}{\mathrm{i}}
\newcommand{\dd}{\mathrm{d}}

\newcommand{\e}[2]{\begin{equation} #1 \label {eq:#2} \end{equation}}

\usepackage{pgfplots}
\usetikzlibrary{plotmarks}
\pgfplotsset{compat=newest}

\makeindex             

\begin{document}

\begin{frontmatter}
\title{A Phase Resonance Approach for Modal Testing of Structures with Nonlinear Dissipation}
\author[addressILA]{Maren Scheel}
\author[addressINM]{Simon Peter}
\author[addressINM]{Remco I.\ Leine}
\author[addressILA]{Malte Krack}

\address[addressILA]{Institute of Aircraft Propulsion Systems, University of Stuttgart, Pfaffenwaldring 6, 70569 Stuttgart, Germany\\ scheel@ila.uni-stuttgart.de, krack@ila.uni-stuttgart.de}
\address[addressINM]{Institute for Nonlinear Mechanics, University of Stuttgart, Pfaffenwaldring 9, 70569 Stuttgart, Germany\\ peter@inm.uni-stuttgart.de, leine@inm.uni-stuttgart.de}

\begin{abstract}
The concept of nonlinear modes is useful for the dynamical characterization of nonlinear mechanical systems. While efficient and broadly applicable methods are now available for the computation of nonlinear modes, nonlinear modal testing is still in its infancy.
The purpose of this work is to overcome its present limitation to conservative nonlinearities.
Our approach relies on the recently extended periodic motion concept, according to which nonlinear modes of damped systems are defined as family of periodic motions induced by an appropriate artificial excitation that compensates the natural dissipation.
The particularly simple experimental implementation with only a single-point, single-frequency, phase resonant forcing is analyzed in detail.
The method permits the experimental extraction of natural frequencies, modal damping ratios and deflection shapes (including harmonics), for each mode of interest, as function of the vibration level.
The accuracy, robustness and current limitations of the method are first demonstrated numerically. The method is then verified experimentally for a friction-damped system. Moreover, a self-contained measure for estimating the quality of the extracted modal properties is investigated.
The primary advantages over alternative vibration testing methods are noise robustness, broad applicability and short measurement duration. The central limitation of the identified modal quantities is that they only characterize the system in the regime near isolated resonances.
\end{abstract}


\end{frontmatter}


\section{Introduction}
The technical relevance of nonlinear vibrations is increasing for a number of reasons: The ever-growing demand for energy and material efficiency leads to lightweight design. Slender structures are more prone to large deformation nonlinearities and lightweight materials commonly exhibit nonlinear behavior. These structures are more likely to be driven into nonlinear self-excited vibrations, as in the case of modern turbine blades where this is a dominant design limitation \cite{Waite.2016}. Finally, novel technologies such as nonlinear vibration absorbers take advantage of nonlinearities to achieve substantially improved performance.

Since the pioneering work of H. Poincar\'{e} in 1905, a solid theoretical framework for nonlinear vibrations has been developed. A versatile toolbox of robust and efficient computational methods for the analysis of nonlinear
vibrations has been established within the last three decades. In contrast, by far most experimental methods are based on linear theory and fail in the presence of nonlinearities. For example, conventional frequency response functions (FRFs) no longer provide a complete picture of the dynamic behavior, since the frequency response depends on the excitation level in the nonlinear case. Moreover, the emergence of new (e.g. sub- or super-harmonic) resonances or the coexistence of steady vibration states have no counterparts in linear theory.

In the following, the experimental methods for the analysis of nonlinear vibrations is briefly addressed. Only the most important methods, in the opinion of the authors, are mentioned, and the presentation is limited to their main idea and current limitations. For a more comprehensive review of methods for nonlinear system identification, we refer to \cite{Kerschen.2006}.

\emph{Nonlinear FRFs} are a straight-forward extension of their linear counterpart. They rely on the measurement of the frequency response to harmonic excitation. In contrast to the linear case, the FRF is determined for different excitation levels of interest. The primary deficiencies of this method are twofold. First, it is practically impossible to realize a purely harmonic excitation due to the interaction between specimen and excitation system \cite{McConnel.2008,Josefsson.2006}, or to keep the level constant throughout resonances. Second, the fine variation of both frequency and level leads to high testing effort and exposing the structure to nonlinear, typically high, vibrations for long times. This can cause severe fatigue and wear damage (\emph{destructive testing}).

The central idea of the \emph{Restoring Force Surface Method} (RFSM) is to determine local nonlinear restoring forces indirectly using the dynamic force equilibrium and measuring mass and deformation quantities \cite{Masri1979,Masri1982,Worden.2001}. An important deficiency of RFSM in its conventional form is that one of the subsystems connected to the nonlinearity must act as a rigid body. Moreover, RFSM is limited to single-valued nonlinear forces, making it unsuitable for hysteretic forces as in the case of dry friction.

The purpose of \emph{nonlinear subspace identification} (NSID) is to derive a nonlinear differential equation system reproducing the response in the tested dynamic regime. The central idea of the method is to interpret the nonlinear forces as internal feedback forces \cite{Adams.1999}. The linear system matrices and coefficients matrices associated to nonlinear forces of an a priori assumed form are then determined by minimizing the deviation between measured and simulated response \cite{Marchesiello2008,noel2013}. An advantage of NSID is that no specific input signal to the system is required, such that typically random broadband excitation signals are used.
This class of excitation signals can also be utilized to identify a polynomial nonlinear state-space model \cite{Paduart2010}, where a linear state-space model is extended by multivariate polynomials and identified based on measured data only. The deficiency is the method's immaturity: The applicability to \eg hysteretic nonlinearities has so far only been demonstrated for a rather idealized numerical example \cite{Noel2017a}.
Perhaps more severely, it appears still to be an unresolved problem to choose the excitation level of a broadband signal for deriving a nonlinear model that is valid for a range of amplitudes of harmonic excitation signals \cite{Noel2017a}. This is especially relevant in case of forced dynamics around a resonance. To enrich the model, several excitation levels could be included in the training data. This would, however, raise the required measurement duration with the aforementioned drawbacks.

The purpose of nonlinear modal analysis is to characterize the dynamical behavior of mechanical systems in terms of natural frequencies, modal damping ratios and vibrational deflection shapes, as a function of the vibration level. These quantities determine at which excitation frequencies resonances are expected, how the resonant vibration energy is distributed within the system, how well the resonances are damped, and how well the system resists dynamic instabilities. The recent literature is particularly rich of contributions on the theoretical framework of nonlinear modes, numerical computation techniques and applications to the analysis and design of nonlinear mechanical systems. On the other hand, studies on experimental nonlinear modal analysis, \ie \emph{Nonlinear Modal Testing} (NMT) are scarce.
Peeters \etal~\cite{Peeters.2011} propose a two-step procedure where they first attempt to \emph{isolate} a nonlinear mode by well-appropriated external forcing, and then to stop the excitation to extract the modal characteristics during the \emph{free decay} (presuming light damping). For the isolation, they suggest to control the force in such a way, that at every excitation point the different harmonics of external force and response displacement have a $90^\circ$ phase lag.
For lightly-damped structures with smooth stiffness nonlinearity, already a single-point, single-harmonic forcing can provide a satisfying isolation of the nonlinear mode if the modes are well-spaced \cite{Peeters.2011,Zapico-Valle2013,Londono.2015,Ehrhardt.2016}.
For systems with nonlinear damping, it is straight-forward to augment this procedure by a damping quantification using time-frequency-analysis of the free decay response \cite{Dion.2013b,Londono.2015,Feldman.1997}, although the term NMT is not used in the literature for these methods.
An important benefit of NMT is the comparatively short measurement duration, since the system's response is only tested once for each vibration level.
The primary deficiency of NMT in its present form is its inaccuracy due to two reasons: First, the transient character of the analyzed response makes the method highly sensitive to noise, which particularly limits the quality of the extracted damping measures. Second, switching-off the excitation usually leads to a finite impulsive loading of the structure. As a consequence, the free decay is spectrally distorted and may contain multiple significant frequency components, so that the extracted damping measure cannot be directly attributed to a specific mode.

The goal of the present work is to extend NMT by a substantially more accurate damping quantification.
The key idea is to analyze stabilized time series under sustained excitation, as opposed to transient free decays, by relying on the recently proposed extended periodic motion definition of damped nonlinear modes \cite{Krack.2015a}.
The experimental method is theoretically derived in \sref{theory}.
To thoroughly assess its accuracy and robustness against noise and imperfect isolation, the method is first applied to a virtual experiment (\sref{numerical}).
In \sref{experimental}, the results of an experimental verification for a friction-damped system are presented.
This paper ends with conclusions and directions of future work (\sref{conclusions}).

\section{Theoretical derivation of the method\label{sec:theory}}
We consider a model of an autonomous mechanical system with $f$ degrees of freedom, of which the dynamics are governed by the ordinary differential equation system
\e{\mbf M\ddot{\bs x} + \mbf K \bs x + \bs g(\bs x,\dot{\bs x}) = \bs 0.}{ode}
Herein, $\bs x \in \mathbb{R}^f$ are generalized coordinates, $\mbf M$, $\mbf K$ are symmetric and positive definite mass and stiffness matrices, respectively, and $\bs g$ are linear and nonlinear damping forces as well as nonlinear restoring forces.
The force term $\bs g$ can represent both local or global nonlinearities. It is assumed that $\bs x=\bs 0$ is an equilibrium of \eref{ode}, \ie $\bs g(\bs 0,\bs 0) = \bs 0$.

The objective of the modal testing method developed in this work is to isolate nonlinear modes in accordance with the \emph{extended periodic motion concept} (\EPMC) proposed in \cite{Krack.2015a}.
The intent of the \EPMC is to design nonlinear modes such that they capture the periodic dynamics of the system described by \eref{ode} under either near-resonant forcing or negative damping of a particular mode.
These dynamic regimes are often of primary technical relevance. For the sake of simplicity, in the following the discussions are limited to nonlinearly damped systems, as these are the subject of interest in the following experimental investigation. However, it should be noted that the extension to self-excited systems is believed to be straightforward.
The motions of \eref{ode} \emph{are made periodic} by introducing an artificial negative damping term $-\xi\mbf M\dot{\bs x}$ that compensates the natural dissipation,
\e{\mbf M\ddot{\bs x} + \mbf K \bs x + \bs g(\bs x,\dot{\bs x}) - \xi\mbf M\dot{\bs x} = 0.}{ode_se}
The family of periodic motions connected to a particular linear mode of vibration are defined as nonlinear modes in accordance with the \EPMC.
The negative damping term does not intend to cancel all the linear and nonlinear damping forces at every location and every time instant. Instead, the purpose of the artificial term is to \emph{compensate the energy} these natural forces dissipate \emph{over a period of vibration}.
The mass proportionality of the term ensures consistency with the linear case under modal damping, where the modes are orthogonal to this term.

This definition is, of course, also consistent with the conventional periodic motion concept for the conservative case, where $\xi=0$.
However when more than one mode of the underlying linear system strongly participates in the dynamics and, at the same time, damping is high, the artificial term may cause distortion.
It was demonstrated for a large range of numerical examples that nonlinear modes in accordance with the \EPMC accurately capture the aforementioned dynamic regime of interest \cite{Krack.2015a,krac2014a,krac2014b,krac2013a}. Previously proposed definitions of nonlinear modes address the damped, as opposed to periodic, dynamics of \eref{ode}, which may lead to comparatively poor representation of vibration properties near resonances and self-excited limit cycles \cite{Krack.2015a}.

\subsection{Mode isolation through force appropriation}
For a given model, the excitation term $\xi\mbf M\dot{\bs x}$ can be simply imposed in a numerical simulation to isolate a particular nonlinear mode.
This corresponds to a forcing applied to all material points.
Such a  globally distributed forcing is practically impossible to achieve in an experiment as an excitation force of the form
\e{\bs f = \xi\mbf M\dot{\bs x}.}{force_appr}
Using a limited number of electrodynamic exciters (shakers) with push rods, for instance, a multi-point excitation can be achieved at best.

Besides the mere force application, another difficulty is to achieve a mass-proportional velocity feedback (self-excitation) in the form $\xi\mbf M\dot{\bs x}$.
Velocity feedback is a well-known method for driving systems into the state of auto-resonance \cite{SOKOLOV.2007,Mojrzisch.2015}. However, this method was so far only applied to systems that can be regarded as single-degree-of-freedom oscillators.
In the present work, a different excitation type is considered: a feedback-controlled forcing.

To isolate a nonlinear mode,
\e{\bs x(t) = \bs x(t+T) = \real { \sum\limits_{n=0}^{\infty} \shpmod_n \ee^{\ii n \ommod t}},}{xoft}
the forcing must have the same periodicity $T=\frac{2\pi}{\ommod}$ as the mode.
Herein, $\tilde{\omega}$ denotes the nonlinear modal frequency, $\ii$ is the imaginary unit, and ${\tilde{\bs{\psi}}}_{n}$ denotes the vector of complex amplitudes of the $n$-th harmonic, \ie ${\tilde{\bs{\psi}}}_{n}$ represents the deflection shape of the respective nonlinear mode and harmonic. For nonlinear systems, this modal deflection shape, as well as the modal frequency, is generally energy (or amplitude) dependent, which is indicated by the $\tilde{()}$-symbol.
In accordance with \eref{force_appr}, the $n$-th force harmonic applied in the direction of the $k$-th coordinate is
\e{F_{k,n} = \ii n\ommod\xi M_{kk} \shpmods_{k,n} + \sum\limits_{j\neq k}^{} \ii n\ommod\xi M_{kj}\shpmods_{j,n}.}{fex}
To achieve this still perfect excitation, the mass matrix $\mbf M=\lbrace M_{ij} \rbrace$ has to be available, and the force at a particular location depends on the magnitudes and phases of the harmonic vibration components at all locations.
These are rather strict requirements, and it is therefore investigated in the following, under which conditions they can be relaxed.

If the mass matrix is diagonal dominant, \ie $\abs{M_{kk}}\gg\abs{M_{kj}}$ for all $j\neq k$, the phase of the force harmonic $F_{k,n}$ applied to coordinate $k$ is mainly determined by the phase of $\shpmods_{k,n}$; \ie the force only has to be in \emph{local phase resonance}. This assumption holds in the case of weak inertia coupling, \eg in the case of a slender structure. This is also the case if the generalized coordinates $\bs x$ are modal coordinates, so that $\mbf M$ is diagonal. However, for this case one would have to be able to apply forcing to specific modes individually. In both cases, the different coordinates have to be driven individually into phase resonance, but there are generally phase lags between different coordinates.
In the special case of a pure standing wave motion, \ie with all material points oscillating phase-synchronously, the phase difference among generalized coordinates is equal to zero and the local phase resonance condition holds as well.

In practice, the number of controllable coordinates and harmonics will be rather limited.
For a multi-point excitation, the mode isolation can theoretically be improved by increasing the number of excitation points. In practice, however, each exciter not only applies a purely external force, but introduces additional elastic and inertia forces. Furthermore, the number of usable shakers is limited by other practical aspects, such as their availability and sufficient space to attach them to the considered specimen.
Similarly, the mode isolation quality improves with the number of successfully appropriated harmonics. Yet, this number is limited, \eg if a feedback-controller is used for enforcing phase resonances, the limitation might stem from the maximum sampling rate or the stability of the controller.
In the remaining part of this work, we explore how well the nonlinear modes can be already isolated, when the external force is applied to only a single coordinate (\ie one direction at a single point), and only the fundamental frequency component of this force is controlled.

\subsection{Single-point, single-frequency force appropriation}
To isolate the periodic motion in a specific nonlinear mode, a forcing $\bs f^{\mathrm{appr}}$ is now considered that is only applied to a single coordinate $k$,
\e{\bs f^{\mathrm{appr}} = \bs e_k \underbrace{\left[\real{\underbrace{\ii\ommod\xi M_{kk}\shpmods_{k,1}}_{F_1}\ee^{\ii\ommod t}} + \real{\sum\limits_{n=2}^{\infty} F_n\ee^{\ii n\ommod t}}\right]}_{f_k^{\mathrm{appr}}}\fk}{fexapprox}
where $\bs e_k$ is the $k$-th unit vector.
Note that the fundamental harmonic force $F_1$ is brought into local phase resonance with the generalized displacement of coordinate $k$.
Higher harmonics $F_n$ of the force with $n>1$ are present in response to the modal motion if no further action is taken.

The phase resonance of the fundamental harmonic force component can be conveniently achieved using a \emph{phase locked loop} (PLL) feedback controller, see \eg \cite{Mojrzisch.2015,Peter.2017}, with the displacement $x_k$, velocity $\dot x_k$ or acceleration $\ddot x_k$ as input signal.
Following this approach, the magnitude $\abs{F_1}$ does not need to be controlled but can in fact be simply recorded for the extraction of the modal properties, as described in \sref{modalproperties}. This is an important advantage, as the mass matrix no longer has to be determined to control the spatial distribution of the force level.
Thus, the practical realization of the appropriated forcing $\bs f_k^{\mathrm{appr}}$ is relatively simple and does not rely on much a priori knowledge. It solely requires the control of its phase while the magnitude is defined by the excitation level.

The appropriated single-point excitation $\bs f^{\mathrm{appr}}$ deviates from the initial self-excitation with $\xi\mbf M\dot{\bs x}$ in the following respects:
\begin{enumerate}[label=(\alph*)]
        \item wrong spatial distribution (local vs. distributed)
        \item uncontrolled higher harmonics (with wrong phase and magnitude)
        \item imperfections introduced by the excitation mechanism, \ie inertia and elastic impedance at the force application point
\end{enumerate}
Of course, (a) and (c) also occur in conventional, \ie linear modal testing using shaker excitation.
In the light of these excitation imperfections, a thorough investigation of the accuracy of the mode quality is required.

A self-contained mode isolation quality indicator is proposed, the power-based mode indicator function (\PBMIF) introduced in \cite{Peter.2017},
\e{\PBMIFs \define \frac{-\activepower}{\apparentpower}.}{pbmif}
Here, $\activepower$ and $\apparentpower$ denote the active and apparent excitation power, respectively,
\e{\activepower = \frac{1}{T}\int\limits_0^T \dot{x}_k f_k^{\mathrm{appr}} \dd t = \real{\sum\limits_{n=1}^{\infty} \frac12 \ii n \ommod \shpmods_{k,n} \overline{F}_{k,n}},}{activePower}
\e{\apparentpower = \sqrt{\frac{1}{T}\int\limits_0^T \dot{x}_k^2 \dd t} \sqrt{\frac{1}{T}\int\limits_0^T \left(f_k^{\mathrm{appr}}\right)^2 \dd t} =  \sqrt{\sum\limits_{n=1}^{\infty}{\frac12 n^2 \ommod^2 \abs{\shpmods_{k,n}}^2}}\sqrt{\sum\limits_{n=1}^{\infty}{\frac12 \abs{F_{k,n}}^2}}, }{}
which can be readily evaluated from the measured force and velocity (or acceleration or displacement) signals at the driving point.
For a single-point, velocity proportional force (\cf \eref{fexapprox}), $\PBMIFs=1$\footnote{It is shown in \aref{power_selfexc} that $\PBMIFs=1$ also holds for the forcing term $\xi\mbf M\dot{\bs x}$.}.
This holds also for a force which is in local phase resonance.
Nonlinear forces cause power transfer among different harmonics, which is captured in $\apparentpower$ but not in $\activepower$, leading to $\PBMIFs<1$.
Thus, the \PBMIF quantifies the combined effect of uncontrolled higher harmonics and inertia or elastic impedance provided by the exciter.
However, the \PBMIF does not indicate the correctness of the spatial force distribution.
It is investigated in this work how the \PBMIF is correlated with the accuracy of the extracted modal characteristics.

\subsection{Extraction of the modal damping ratio\label{sec:modalproperties}}
When the excitation force is successfully appropriated, the motion of the system becomes periodic and takes the form given in \eref{xoft}.
The modal frequency $\ommod$ and modal deflection shape harmonics $\shpmod_0,\shpmod_1,\shpmod_2,\ldots$ can thus be directly recorded.
The modal damping ratio $\Dmod$ is estimated by the balance between provided and dissipated power. Since only the fundamental harmonic is controlled, the balance is limited to the associated component.

In the nonlinear mode, the active power of the artificial negative damping term can be written in terms of nonlinear modal coordinates as
\begin{equation}\label{PowerDissMode}
P  = \sum_{n=1}^{\infty} \dfrac{1}{2} \left( n \tilde{\omega} \right)^2 \tilde{\bs{\psi}}_n^{\rm H}  \xi \mbf{M} \tilde{\bs{\psi}}_n = \sum_{n=1}^{\infty} \dfrac{1}{2} \left( n \tilde{\omega} \right)^2 q^2 \tilde{\bs{\phi}}_n^{\rm H}  \xi \mbf{M} \tilde{\bs{\phi}}_n .
\end{equation}
The mode shape harmonics $\shpmodnorm_n$ are normalized such that the modal mass of the fundamental harmonic is unity; \ie $\shpmodnorm_1^{\mathrm{H}}\mbf M \shpmodnorm_1 = 1$ and $\shpmod_i = q \shpmodnorm_i$.
The self-excitation factor $\xi$ is, per definition, the same for all harmonics $n$, such that it can be identified, for instance from the first harmonic component of the active power, $P_1$ for $n = 1$, as
\begin{equation}
\xi = \dfrac{2 P_1}{\tilde{\omega}^2 q^2 \shpmodnorm_1^{\mathrm{H}}\mbf M \shpmodnorm_1} = \dfrac{2 P_1}{\tilde{\omega}^2 q^2}.\label{eq:xi_activepower}
\end{equation}

For a forced system, the first harmonic component of the active power $P_1$ provided by the excitation force $\bs f_k^{\mathrm{appr}}$  (see \eref{activePower}) is inserted in Eq.\ \eqref{eq:xi_activepower} and the modal damping ration $\Dmod$ is computed from the self-excitation factor $\xi$ as
\e{\Dmod =   \dfrac{\xi}{ 2 \ommod}
}{dmod}
to be fully consistent with the linear case.

The required mass matrix is proposed to be estimated as
\e{\mbf M \approx \mbf M_{\mathrm{exp}} \define \left(\bs\Phi\tra\right)^+~\left(\bs\Phi\right)^+\fk}{mexp}
with the mass-normalized mode shapes $\bs\Phi = \matrix{ccc}{\bs\phi_1 & \bs\phi_2 & \cdots}$ obtained from linear modal testing. Note that the $()^+$ operator denotes the generalized inverse for cases in which the eigenvector matrix is non-square, \eg when more points are measured than modes are estimated.

\subsection{Overview of the nonlinear modal testing method}
The overall nonlinear modal testing method is summarized in \fref{testing_procedure}. Standard vibration testing equipment is used to apply a single-point forcing to the specimen and measure the response.
Conventional linear modal testing is first applied, where the vibration level should remain sufficiently small so that dynamics can be regarded as linear.
A specific mode has to be selected for the nonlinear testing and excitation levels have to be provided which are determined by the input voltage to the shaker.
A PLL phase controller is used to adjust the excitation frequency until the fundamental harmonic forcing is in phase resonance with the response at the driving point. The schematic of the PLL controller, including the transfer functions used in this work, is provided in \aref{pll}.
The excitation frequency is an internal variable of the PLL and can be directly recorded once the controller reaches a locked state.
This facilitates avoiding leakage or windowing-induced distortions, as the discrete Fourier transform can be applied directly to a periodic time frame of the steady-state signals.
The larger the number of considered periods, the longer is the testing duration, and the better can the inevitable noise be averaged out.
This way, the modal properties are obtained for each excitation level.
Although not mandatory, the excitation level was in this study started from low level and then increased. The lowest level should be high enough to permit stable operation of the controller under the possible effect of noise. The highest level should, of course, be well below the operating limits of the testing equipment and the strength limit of the specimen.
\begin{figure}[h!]
	\centering
	\begin{subfigure}[]{0.39\textwidth}
		\centering
		\includegraphics[width=\textwidth]{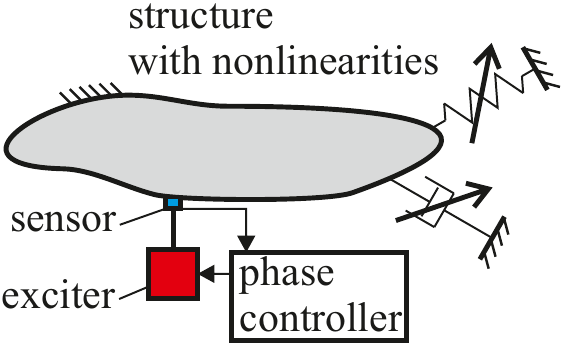}
		\caption{}
	\end{subfigure}\hfill
	\begin{subfigure}[]{0.59\textwidth}
		\centering
		\def\svgwidth{\textwidth}
		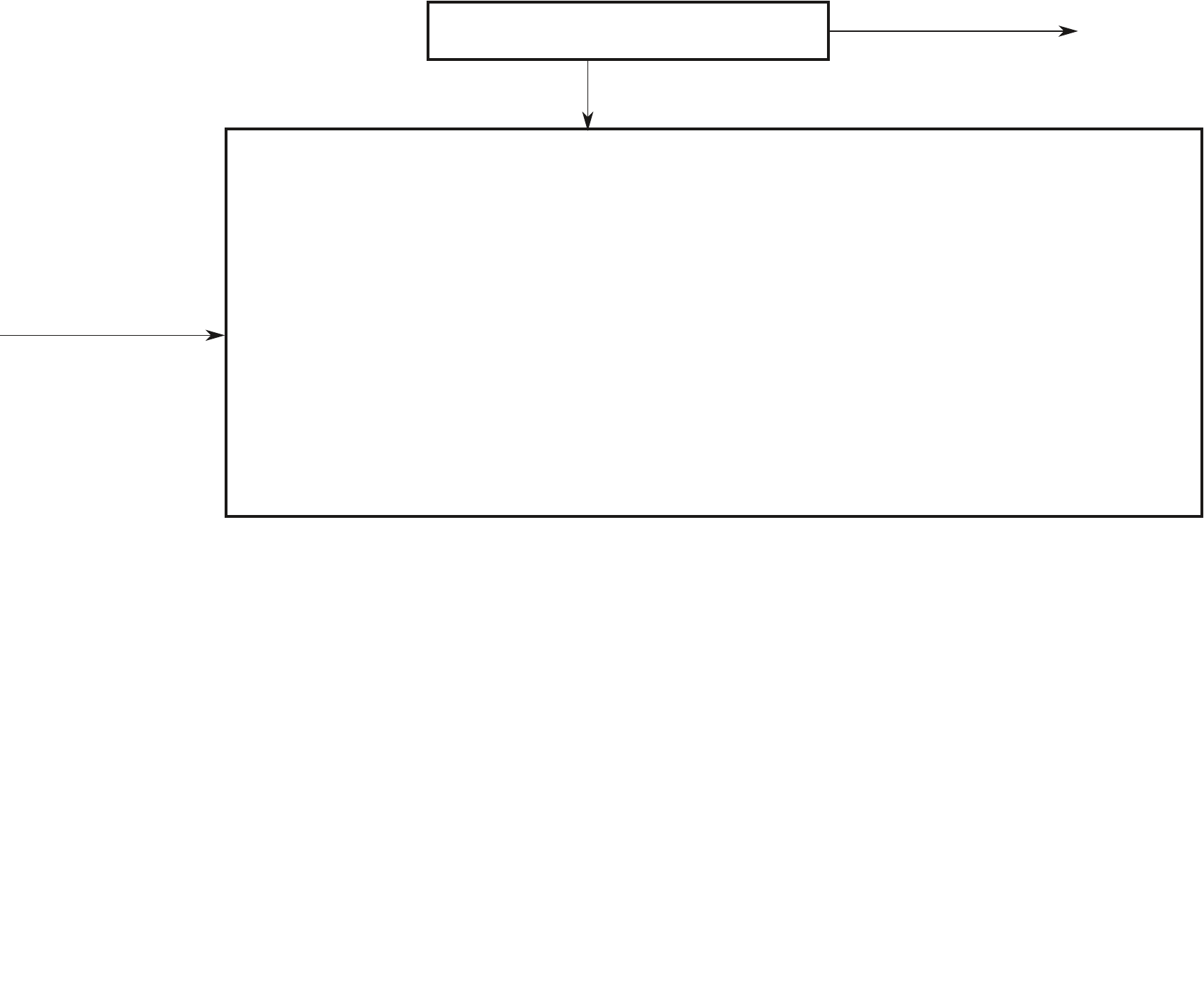\caption{}
	\end{subfigure}
	\caption{Overview of the nonlinear modal testing method: (a) instrumentation of the specimen, (b) testing procedure.}
	\label{fig:testing_procedure}
\end{figure}

\subsection{Comparison to modal testing of the underlying conservative system}
The by far largest amount of work related to nonlinear modes focused on conservative nonlinear systems.
In this mindset, the typically weak damping present in reality is regarded as parasitic, and the goal of modal testing is in this case to identify the underlying conservative system.
To this end, the parasitic damping can be compensated by appropriate forcing.
This force appropriation can, for instance, be achieved using phase resonant testing \cite{Peeters.2011,Ehrhardt.2016,Peter.2018}, \eg using PLL control \cite{Peter.2018}.
As a consequence, the testing procedure can be exactly the same as that proposed in the present work for the single-point excitation case.
The conceptual difference is that this type of forcing was derived as substitute for the self-excitation with $\xi\mbf M\dot{\bs x}$ in accordance with the \EPMC in our work.
Using the \EPMC permits the extraction of a modal damping ratio of nonlinearly damped systems.

\section{Numerical assessment of validity and robustness\label{sec:numerical}}
In this section, it is investigated how accurate the proposed method is in the light of the imperfections induced by the single-point single-frequency force appropriation.
To this end, the impedance of the exciter is included in the model, and the effect of its placement, measurement noise and erroneous identification of the underlying linear modes is studied.
To have better control over these imperfections, the testing procedure is simulated (\emph{virtual experiment}). An experimental verification is provided in \sref{experimental}.
As reference, the direct computational nonlinear modal analysis is used. To this end, Harmonic Balance is applied to \eref{ode_se}, considering the first five harmonics. Inclusion of 20 harmonics did not influence the modal characteristics significantly. Furthermore, no indications of internal resonances, such as turning points or other irregularities were found for more harmonics. For a more detailed description of this method, we refer to \cite{Krack.2015a}.

The model of specimen and instrumentation is specified in Fig.\ \ref{fig:num_beam} and Tab.\ \ref{tab:num_beam}.
A cantilevered friction-damped beam serves as specimen. It is described by seven linear Euler-Bernoulli beam elements and a nonlinear elastic Coulomb friction element. Additional light damping forces $\beta \mbf{K}\dot{\bs x}$ are introduced such that the lowest frequency modes of the underlying linear model (sticking friction contact) have the natural frequencies and modal damping ratios as specified in \tref{num_modal}. The virtual experiment focusses on the system's lowest-frequency bending mode.
The model of the electrodynamic excitation system contains the stinger stiffness $k_{\text{Stinger}}$, coil and shaker table mass $m_C$ and $m_{\rm{T}}$, respectively, and the electric circuit connected to the controller. The electrodynamic force on the coil acting is $Gi$, where $i$ is the electrical current and $G$ is the force generating constant. The parameters of the shaker model are listed in Tab.\ \ref{tab:num_beam} and represent the Br\"uel and Kj\ae r Vibration Exciter 4808 as identified in \cite{Morlock2015}.
The phase resonance of the fundamental harmonic forcing is enforced using a PLL controller with the properties listed in \ref{append:pll}. After a waiting time of about 11 seconds, the controller reaches a locked state and 200 excitation periods are recorded.
The modal frequency can be directly extracted from the PLL controller, and the required harmonics of the forcing and the deflection variables are determined by discrete Fourier transform. The excitation force is varied between 0.1 N and 3 N.
Four sensors measure the lateral displacement of points two, four, five and six.
The first two modes of the underlying linear system (sticking friction contact) are numerically determined and used for the estimation of the mass matrix (\cf Eq.\ \eqref{eq:mexp}).

\begin{table}
	\begin{minipage}[b]{0.6\textwidth}
		\centering
		\def\svgwidth{0.8\textwidth}
		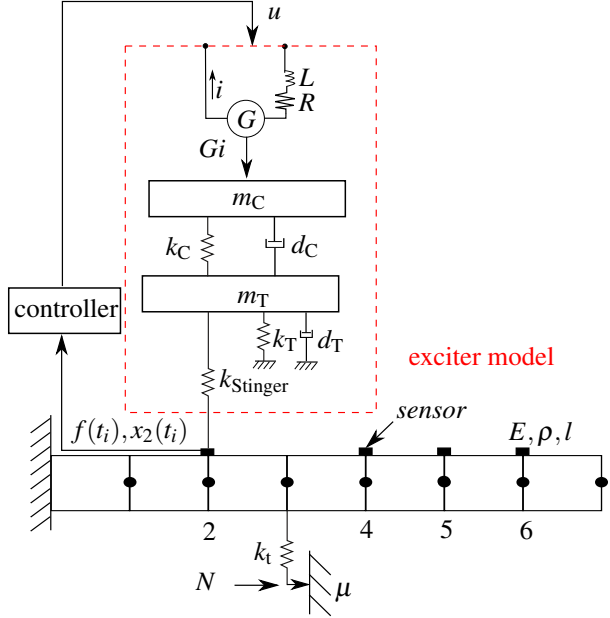
		\captionof{figure}{Model of specimen and instrumentation for the virtual experiment.}
		\label{fig:num_beam}
	\end{minipage}
\hfill	
	\begin{minipage}[b]{0.35\textwidth}
		\centering
		\begin{tabular}{l|l}
			\hline
			Parameter & Value\\
			\hline
	 		$E$ & 185 GPa\\
	 		$\rho$ & $7830 \text{ kg}/\text{m}^3$\\
			$l$ & $0.7 \text{ m}$\\
	 		$k_{\rm{t}}$ & 8000 N/m\\
	 		$\mu N$ & 1 N \\
	 		$\beta$ & $10^{-4}$ s\\
	 		$m_{\rm{T}}$ & 0.024 kg\\
	 		$m_C$ & 0.019 kg\\
	 		$k_{\rm{T}}$ & 20707 N/m\\
	 		$k_{\rm{C}}$ & $8.42\cdot 10^{7}$ N/m \\
	 		$d_{\rm{T}}$ & $28.33 $ Ns/m\\
	 		$d_{\rm{C}}$ & $57.17 $ Ns/m\\
	 		$G$ & $15.48 $ N/A\\
	 		$L$ & $140 \cdot 10^{-6}$ H\\
	 		$R$ & $3 \text{ }\Omega$\\
	 		$k_{\text{Stinger}}$ & $1.32\cdot 10^{14} $ N/m\\
			\hline
		\end{tabular}
		\captionof{table}{Parameters of specimen and instrumentation for the virtual experiment.}\label{tab:num_beam}
	\end{minipage}
\end{table}
\begin{table}[tb]
	\centering
	\begin{tabular}{l|ccc}
		\hline
		linear mode & 1& 2& 3\\
		\hline
		$\omega$ in rad/s& 139.5& 760.9&  2052.7\\
		\hline
		$\delta$ in \% & 0.6 &  3.7 & 10.2\\
		\hline
	\end{tabular}
	\caption{Lowest modal frequencies and associated damping ratios of the linearized specimen model.\label{tab:num_modal}}
\end{table}

\subsection{Influence of the exciter location}
\begin{figure}
	\centering
	\begin{subfigure}[]{0.49\textwidth}
		\centering
		\includegraphics{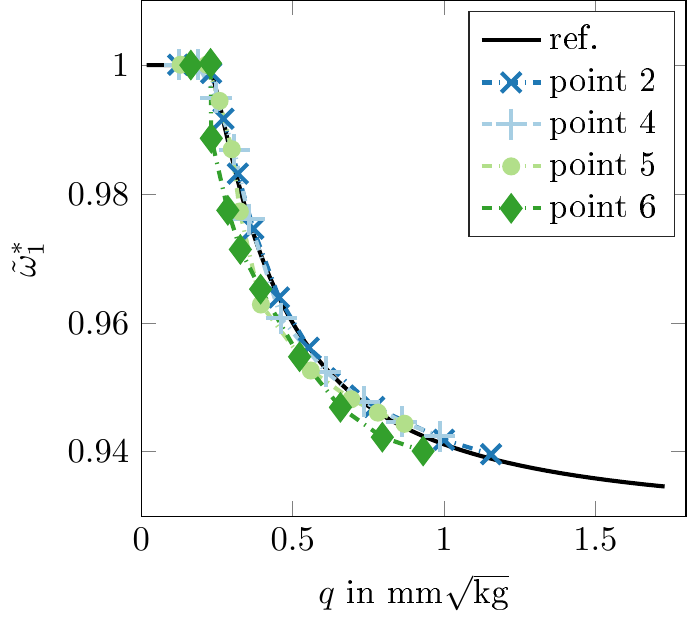}
		\caption{}
	\end{subfigure}\hfill
	\begin{subfigure}[]{0.49\textwidth}
		\centering
		\includegraphics{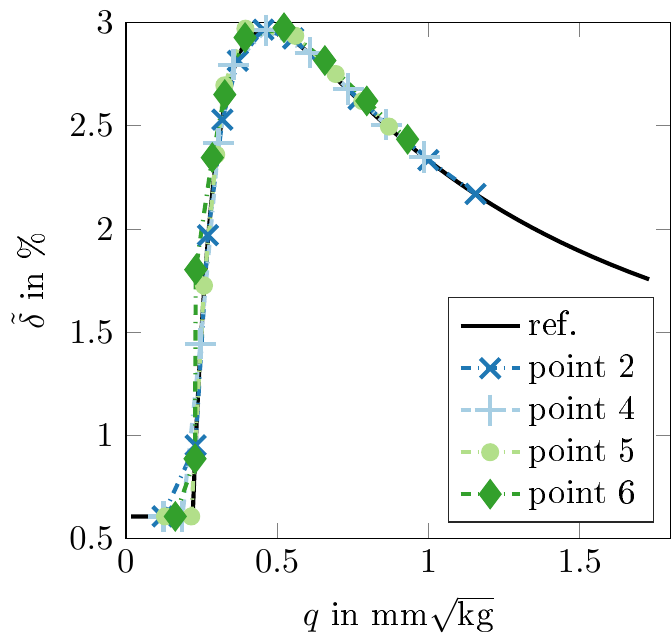}\caption{}
	\end{subfigure}
	\caption{Modal properties of the first nonlinear mode vs. vibration level extracted for different excitation points: (a) natural frequency (b) damping ratio.}
	\label{fig:points2456}
\end{figure}

First, the influence of the exciter location on the mode isolation quality is investigated.
Fig.\ \ref{fig:points2456} shows the extracted modal frequencies, normalized by the linear modal frequency, $\tilde{\omega}_{1}^\ast = \tilde{\omega}_{1}/\omega_{1}$ and modal damping ratios $\tilde{\delta}$ as function of the modal amplitude $q$.
The results agree well with the numerical reference.
When the exciter is attached to points five or six, the accuracy is slightly lower but still regarded as satisfying.
Apparently, the single-point single-frequency force appropriation is in this case sufficient to isolate the nonlinear mode.

\begin{figure}
	\centering
	\begin{subfigure}[]{0.49\textwidth}
		\centering
		\includegraphics{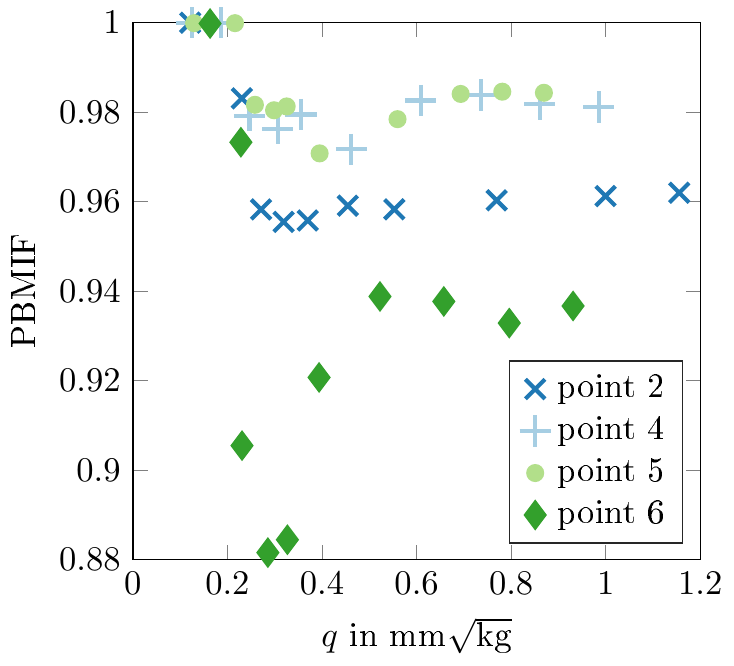}\caption{}\label{fig:PBMIF_q}
	\end{subfigure}
	\begin{subfigure}[]{0.49\textwidth}
		\centering
		\includegraphics{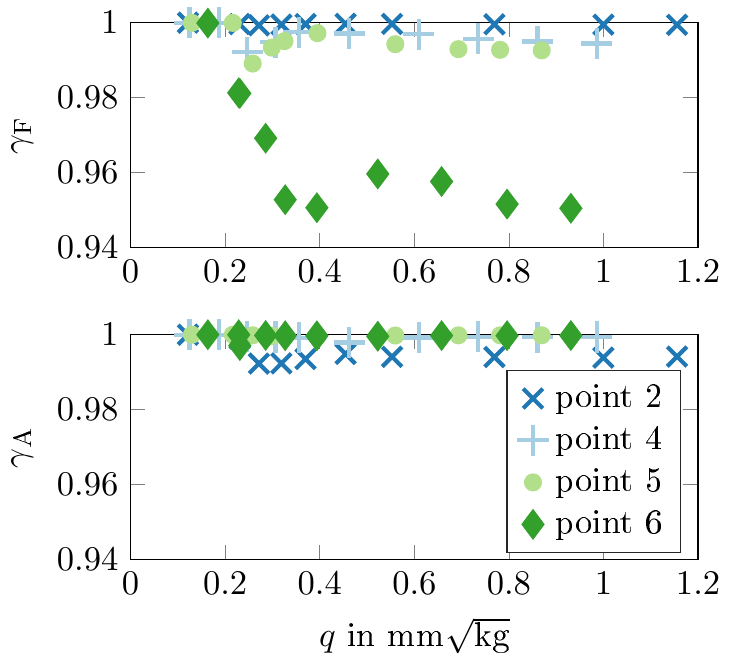}\caption{}\label{fig:gamma_q}
	\end{subfigure}
	\caption{Force appropriation measures vs. vibration level for different excitation points: (a) \PBMIF, (b) fundamental harmonic content of force $\gamma_\text{F}$ and acceleration $\gamma_\text{A}$.}	
\end{figure}

The variation of the modal deflection shape is illustrated in \fref{modal_amp} in terms of the modal participation factor $\bs{\Gamma} = \bs{\Phi} ^+ \tilde{{\bs{\phi}}}_{1}$ of the fundamental harmonic component $\tilde{{\bs{\phi}}}_{1}$. As the system is damped, its coordinates are generally not in phase and $\tilde{{\bs{\phi}}}_{1}$ (and thus $\bs{\Gamma}$) is complex-valued. The contribution of the second mode (and higher modes) is negligible; \ie the mode shape does not considerably change with the vibration level in this case.
For both excitation point two and six, the phase of the first linear mode is close to zero such that the motion is close to synchronous. The phase of the second linear mode is not meaningful as it does not participate considerably.
\begin{figure}
	\centering
	\begin{subfigure}[]{0.49\textwidth}
		\centering
		\includegraphics{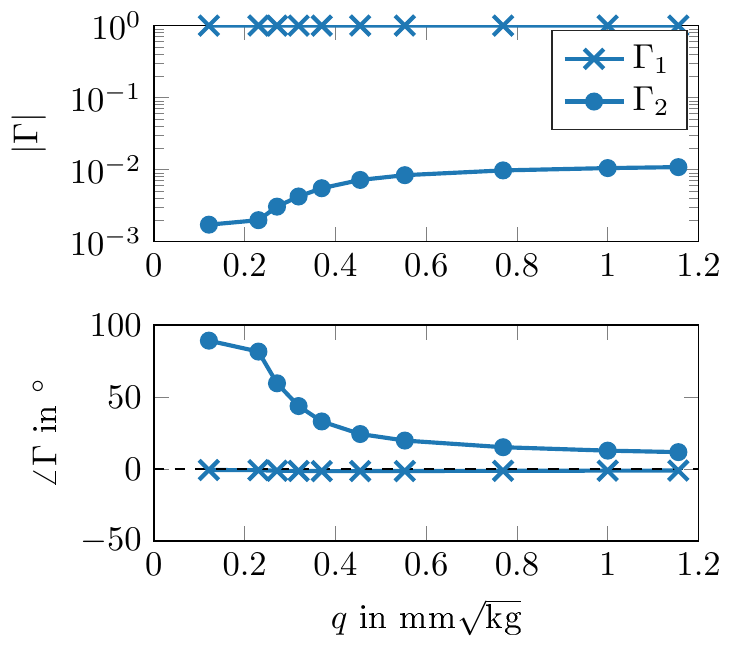}\caption{}
	\end{subfigure}
	\begin{subfigure}[]{0.49\textwidth}
		\centering
		\includegraphics{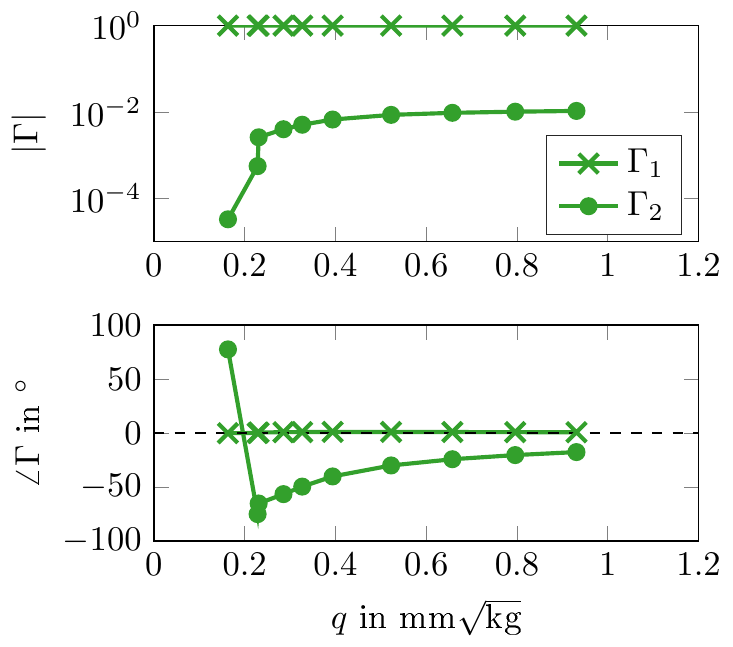}\caption{}
	\end{subfigure}
\caption{Participation of the underlying linear modes in the first nonlinear mode vs.\ vibration level: (a) excitation at point two, (b) excitation at point six.}
\label{fig:modal_amp}
\end{figure}
The \PBMIF is illustrated in Fig.\ \ref{fig:PBMIF_q} as a function of the vibration level.
Consistent with the comparatively low accuracy of the extracted modal properties for excitation point six, the \PBMIF deviates considerably from its ideal value of unity.
On the other hand, the \PBMIF yields lower values for excitation point two than for point four or five, which contradicts the higher accuracy for point two.
Hence, further investigations are required to better understand the correlation between \PBMIF and accuracy of the extracted modal characteristics.

The fundamental harmonic content of force, $\gamma_\text{F}$, and acceleration at the load application point, $\gamma_\text{A}$, is depicted in Fig.\ \ref{fig:gamma_q}.
Apparently, higher harmonics are more pronounced in the excitation force when the shaker is placed closer to the free end of the beam, which is consistent with observations in previous studies, see \eg \cite{Schwingshackl.2014}.

\subsection{Robustness against measurement noise}
In the following, the system is excited at point two unless otherwise stated.
To assess sensitivity against noise, inevitably encountered in reality, band-limited white noise was added to the force and displacement signals. The correlation time was set to 0.01 ms, and two different noise levels are investigated whose power spectral densities (PSD) are specified in Tab.\ \ref{tab:noiseSettings}. The resulting signal to noise ratio (SNR) is also given in the table for the force signal and displacement at the excitation point. As the noise level is assumed to be independent of the vibration level, the SNR improves for higher vibration levels.

\begin{figure}
	\centering
	\begin{subfigure}[]{0.49\textwidth}
		\centering
		\includegraphics{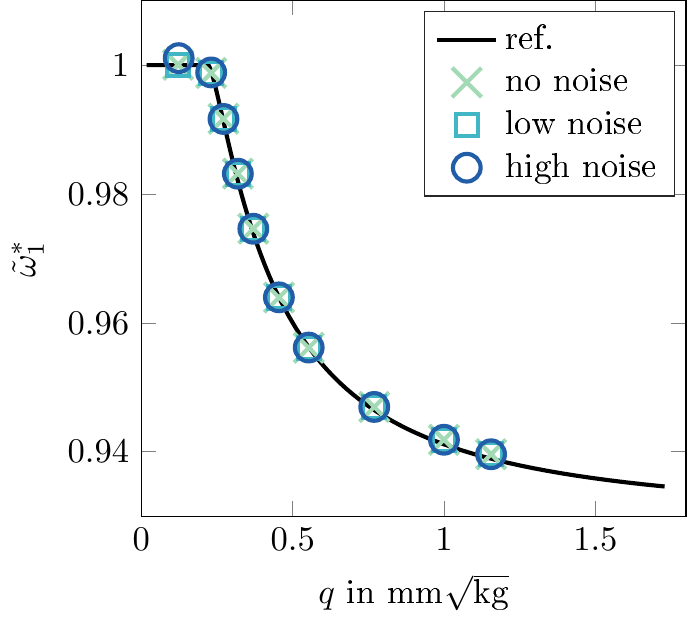}\caption{}
	\end{subfigure}
	\begin{subfigure}[]{0.49\textwidth}
		\centering
		\includegraphics{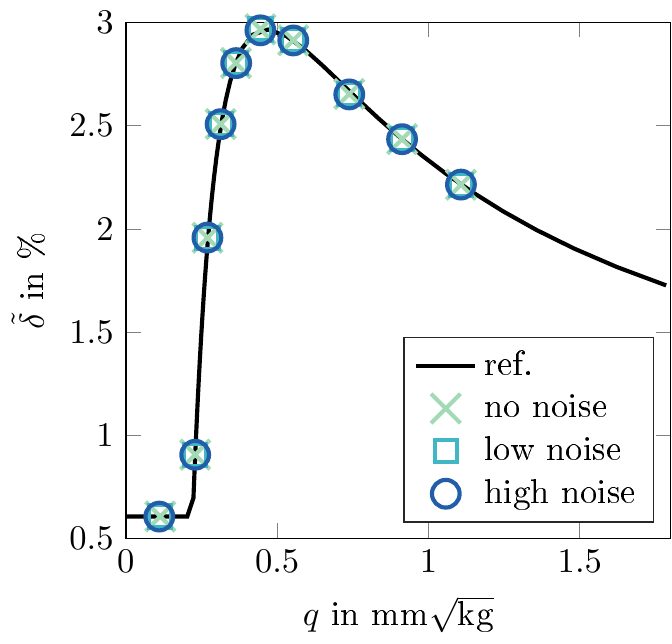}\caption{}
	\end{subfigure}
	\caption{Effect of measurement noise on the modal characteristics for excitation point two: (a) natural frequency, (b) modal damping ratio.}\label{fig:noise}
\end{figure}

\begin{table}
		\centering
		\begin{tabular}{l c|c c}
	\multicolumn{4}{l}{\textbf{force}}\\		
	\hline
	noise level & & low & high\\
	\hline &  &  & \\[-8pt]
 	PSD & & $2 \cdot 10^{-9}$ W/Hz & $2 \cdot 10^{-8}$  W/Hz\\
	SNR & \begin{tabular}{c} low vibr. levels \\ high vibr. levels \end{tabular} & \begin{tabular}{c} 14.5 dB \\ 39.1 dB \end{tabular}  & \begin{tabular}{c} 4.8 dB \\ 33.0 dB \end{tabular}\\
	\hline
		\multicolumn{4}{l}{\textbf{displacement at excitation point}}\\		
	\hline
	noise level & & low & high\\
	\hline &  &  & \\[-8pt]
	PSD & & $4.5 \cdot 10^{-16}$ W/Hz & $4.5 \cdot 10^{-15}$  W/Hz\\
	SNR & \begin{tabular}{c} low vibr. levels \\ high vibr. levels \end{tabular} & \begin{tabular}{c} 11.2 dB \\ 31.2 dB \end{tabular}  & \begin{tabular}{c} 0.8 dB \\ 20.7 dB \end{tabular}\\
	\hline
\end{tabular}
\caption{PSD and SNR of the measurement noise (virtual experiment)}\label{tab:noiseSettings}
\end{table}
The modal characteristics extracted without and with different levels of noise are depicted in Fig.\ \ref{fig:noise}.
The results are highly robust against the considered noise levels.
This indicates that the 200 steady-state excitation periods recorded for modal property extraction are sufficient to average out the random noise.
For noise levels with PSD of one order of magnitude larger than those listed in \tref{noiseSettings}, the noise level in the excitation location displacement exceeds the signal level, yielding negative SNR. For noise levels another order of magnitude larger, it was found that the PLL controller did not converge to a locked state. Such a failure of the controller is more likely to occur in the linear regime, where vibration levels and, thus, SNRs are low.

\subsection{Robustness against erroneous identification of the underlying linear modes}
The proposed method relies on the mass normalized mode shapes of the underlying linear system. These mode shapes are used to estimate the mass matrix in order to determine the modal amplitude and the modal damping ratio.
For the virtual experiment, the numerically exact mode shapes were so far adopted. Wrong scaling of the modal matrix $\bs\Phi$ directly affects the estimated modal damping ratio. Given a scaled modal matrix $\kappa \bs\Phi$, the modal amplitude $q$ is reduced by 1/$\kappa$ which multiplies the modal damping ratio $\Dmod$ by $\kappa^2$ (see \eref{xi_activepower}). For an error of 10 \%, \ie $\kappa = 1.1$ or $\kappa = 0.9$, the error of the modal damping ratio is 21 \% and 19 \%, respectively.

\begin{figure}
	\centering
	\begin{subfigure}[]{0.49\textwidth}
		\centering
		\includegraphics{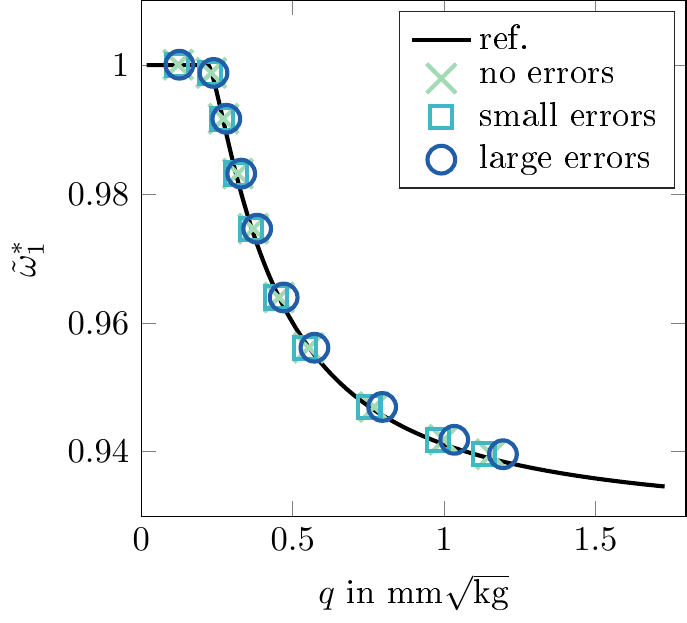}\caption{}
	\end{subfigure}
	\begin{subfigure}[]{0.49\textwidth}
		\centering
		\includegraphics{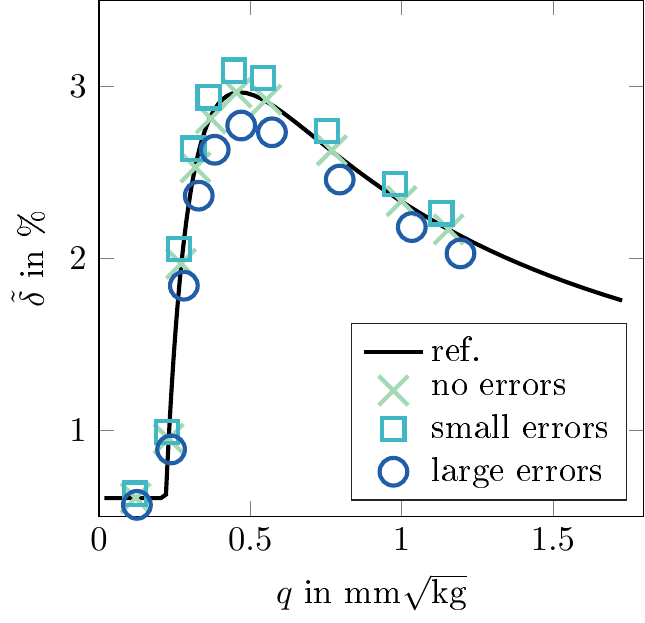}\caption{}
	\end{subfigure}
	\caption{Effect of erroneous linear mode shapes on the modal characteristics for excitation point two: (a) natural frequency, (b) modal damping ratio.}\label{fig:modal errors}
\end{figure}

Now, normally distributed random errors are added with a level of 5~\% and 10~\% of the norm of the undistorted modal matrix $\bs\Phi$.
The effect of these errors on the extracted modal characteristics is illustrated in Fig.\ \ref{fig:modal errors}. Note that these errors do not affect the isolated nonlinear mode, but only the modal property extraction from the recorded time series. Accordingly, the natural frequencies are correctly measured, but the modal amplitude axis is erroneously stretched. Both the modal damping ratio and modal amplitude axes are erroneously stretched depending on the mode shape error. Apparently, the method is quite robust against noisy mode shapes, as the relative errors of the modal characteristics are much smaller than the errors imposed on the mode shapes.

\subsection{Increasing influence of nonlinearity on the system}
\begin{figure}
	\centering
	\begin{subfigure}[]{0.49\textwidth}
		\centering
		\includegraphics{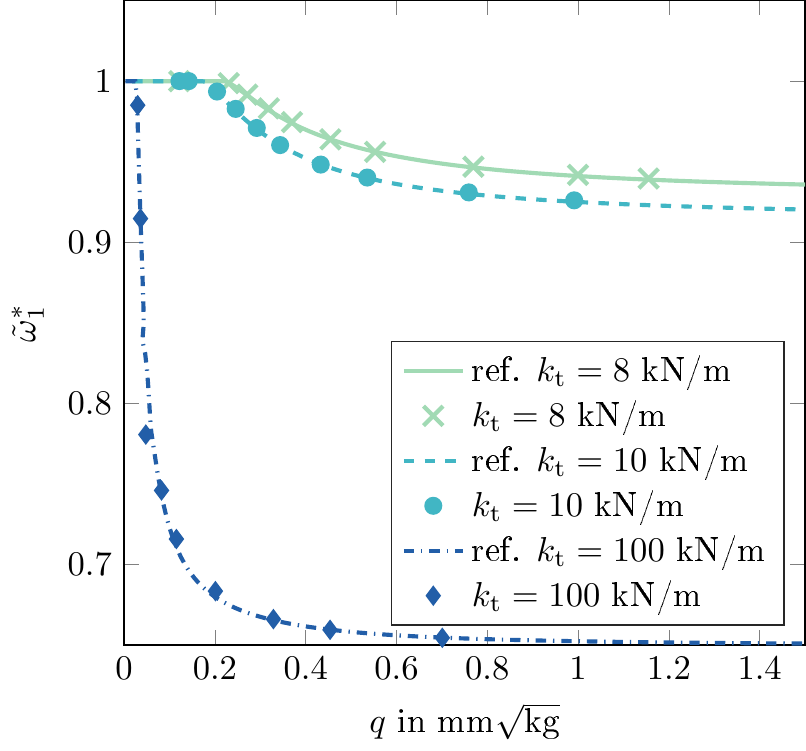}
		\caption{}\label{fig:kt2_eig}
	\end{subfigure}\hfill
	\begin{subfigure}[]{0.49\textwidth}
		\centering
		\includegraphics{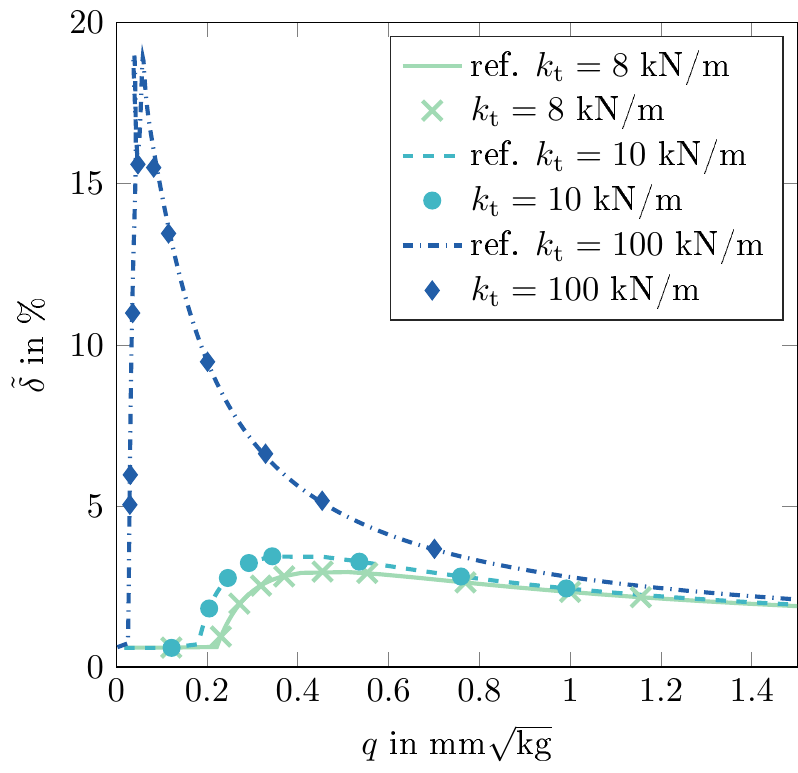}\caption{}
		\label{fig:kt2_damp}
	\end{subfigure}
	\begin{subfigure}[]{0.49\textwidth}
		\centering
		\includegraphics{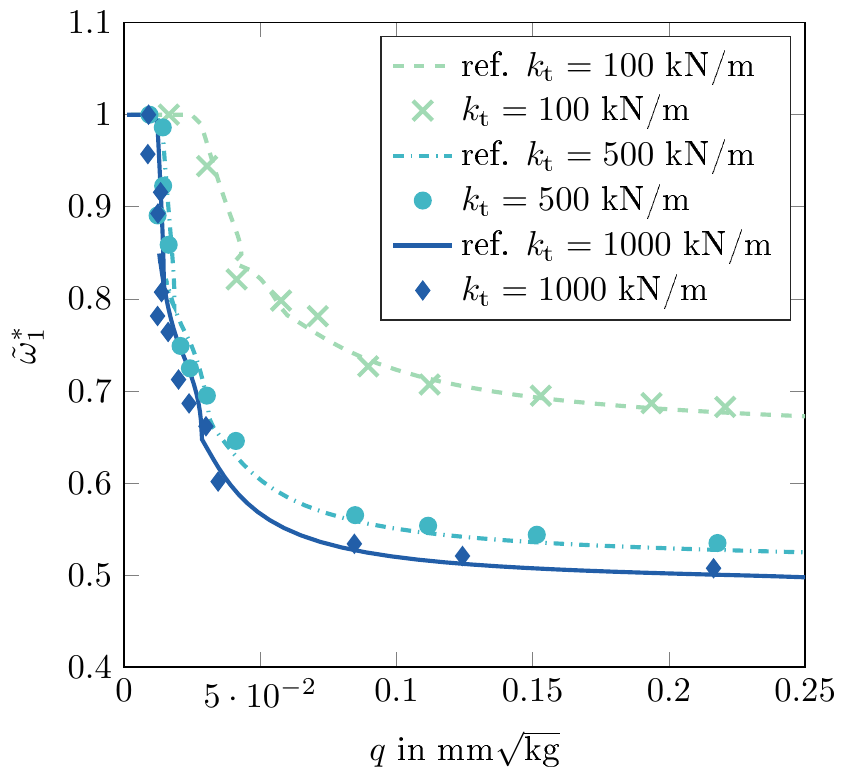}\caption{}
		\label{fig:kt6_eig}
	\end{subfigure}\hfill
	\begin{subfigure}[]{0.49\textwidth}
		\centering
		\includegraphics{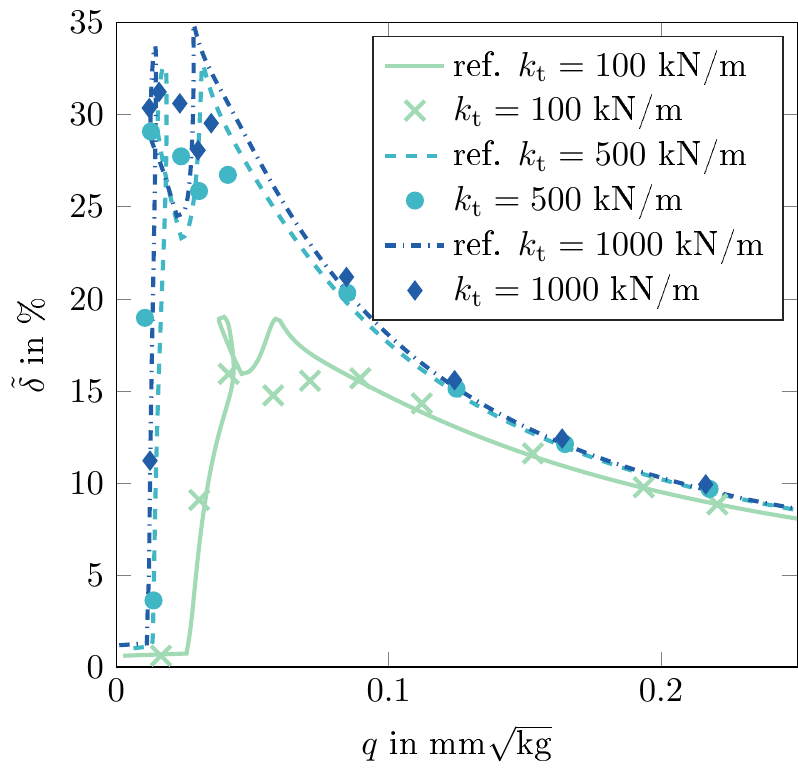}\caption{}
		\label{fig:kt6_damp}
	\end{subfigure}
	\caption{Effect of stiffness $k_{\rm{t}}$ of nonlinear friction element on the modal characteristics: (a-b) excitation at point two for low and moderate $k_{\rm{t}}$, (c-d) excitation at point two for high $k_{\rm{t}}$.}
\end{figure}

In the foregoing studies, both modal frequencies and damping ratios vary only moderately with the vibration level. Moreover, the deflection shape does not change significantly. To explore the limits of utility of the proposed approach, the tangential stiffness $k_{\rm{t}}$ is increased to magnify these nonlinear variations. Exciting at point two, the modal frequencies and damping ratios can be extracted with satisfying precision up to $k_{\rm{t}} = 100 \text{ kN/m}$ (see Fig.\ \ref{fig:kt2_eig} and \ref{fig:kt2_damp}), even for damping ratios as high as 15\% and frequency shifts of 35 \%.

\begin{figure}[tb]
	\centering
	\includegraphics{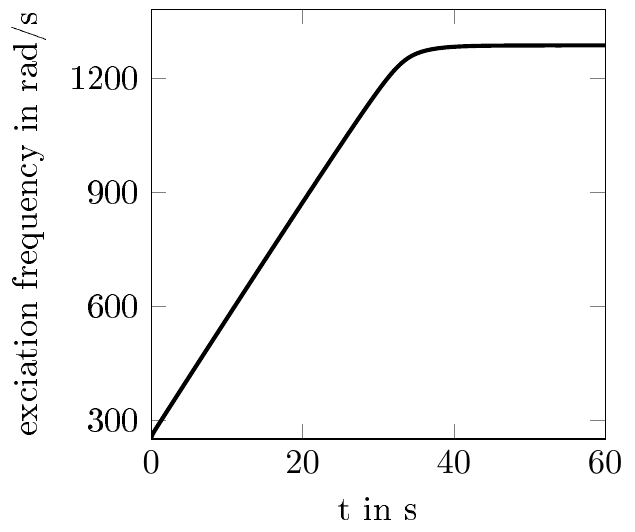}
	\caption{Locking of the PLL on the second mode for force appropriation at point two, $k_t = 500$ kN/m.}\label{fig:high_kt_pll}
\end{figure}

For higher $k_{\rm{t}}$ the controller locked onto the second rather than the first mode in some cases. For instance, this occurred when the exciter was applied at point two (see Fig.\ \ref{fig:high_kt_pll}) where the lateral deflection is relatively high for the second mode shape as compared to the first mode shape (with high $k_{\rm{t}}$ attached to point three).
As a consequence, the first nonlinear mode could not be isolated with forcing applied to point two, even when the initial excitation frequency was set to the linear one.
However, the first nonlinear mode could be isolated also for higher $k_{\rm{t}}$, when forcing was applied to point six, see Fig.\ \ref{fig:kt6_eig} and \ref{fig:kt6_damp}.
Yet, the precision of the extracted modal damping ratios suffers in the high damping regime.

Concluding the virtual experiment, the proposed modal testing method is capable of isolating nonlinear modes under strongly nonlinear stiffness and damping effects. The method demonstrated high robustness with regard to shaker placement, measurement noise and erroneous identification of the underlying linear mode shapes.
So far, the range of utility of the method appears to be largely limited by the operating range of the controller, and enhancements might be required for specific nonlinearities.

\section{Experimental verification for a friction-damped system\label{sec:experimental}}
\begin{figure}
	\centering
	\def\svgwidth{0.8\textwidth}
	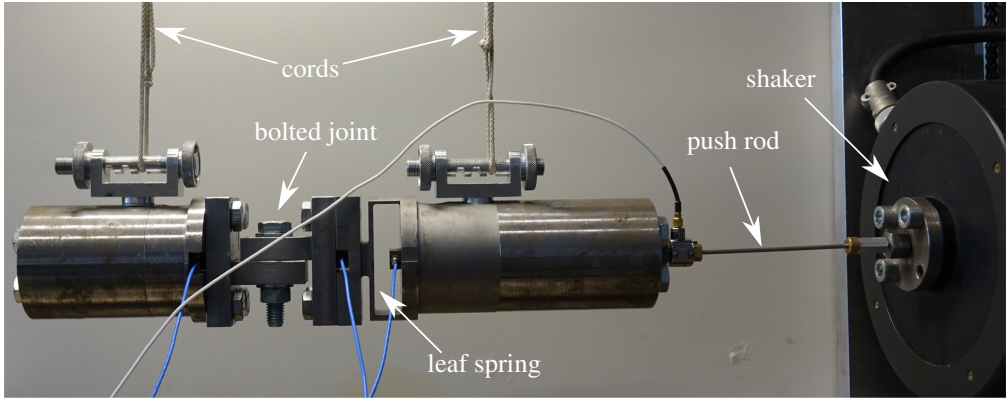
	\caption{Photo of the joint resonator used for experimental verification.}
	\label{fig:test_rig}
\end{figure}

\begin{figure} [tb]
	\center
	\includegraphics{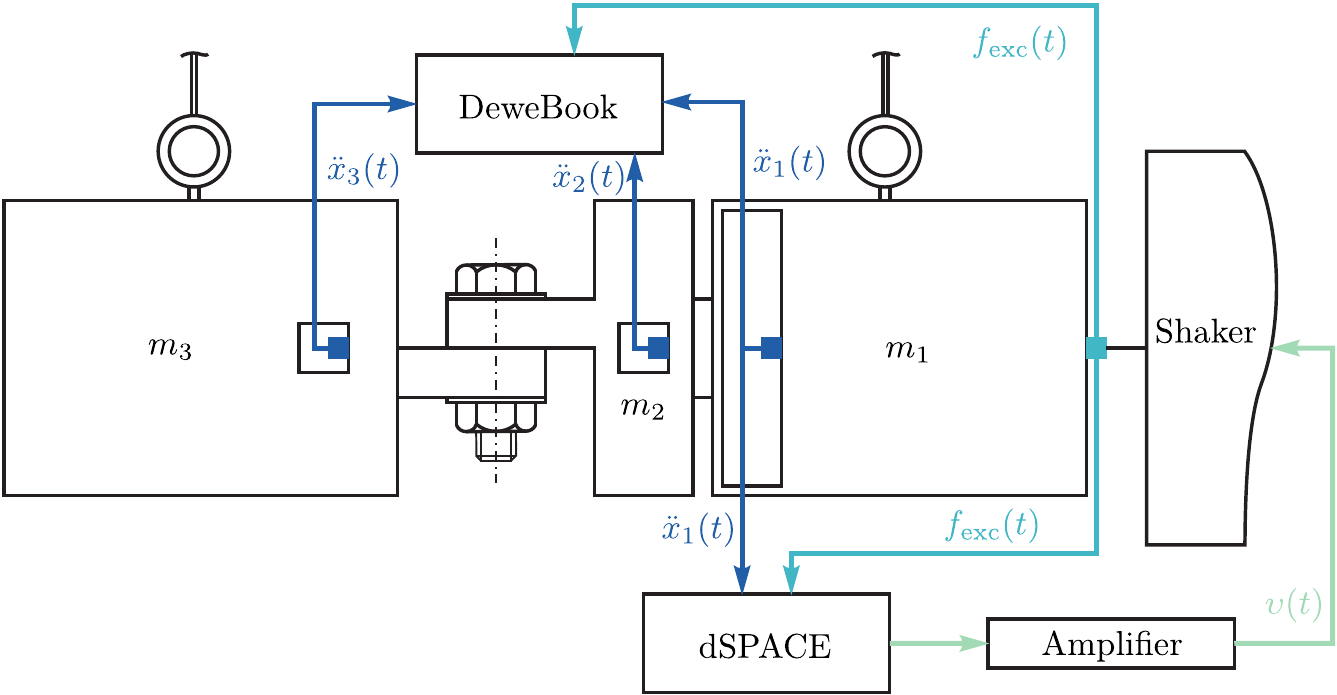}
	\caption{Schematic of the specimen and instrumentation.}
	\label{fig:exp_setup}
\end{figure}

\begin{figure}

\centering
\includegraphics[width=0.4\textwidth]{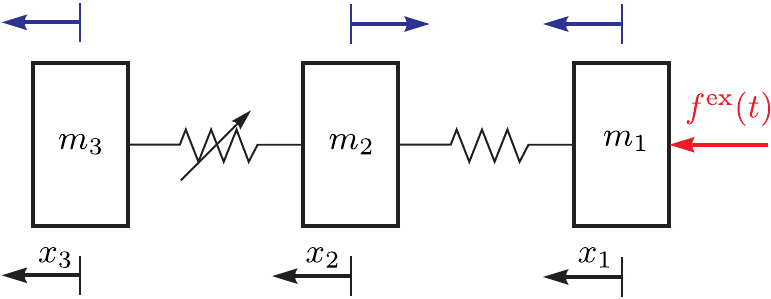}
	\caption{Three-degree-of-freedom model of the joint resonator. The shape of the mode of interest is indicated by arrows.}
	\label{fig:minimal_model}

\end{figure}

The proposed nonlinear modal testing method was applied to the \emph{joint resonator} depicted in Fig.\ \ref{fig:test_rig} and \ref{fig:exp_setup}.
The specimen is a known benchmark system for the dynamic characterization of bolted joints \cite{bohl1987, Gaul1997,Segalman2009,Suess2016}. It is known from previous studies that its low-frequency dynamics can be well-described by a chain of three lumped masses \cite{Suess2016,Ehrlich2016}, connected as illustrated in Fig.\ \ref{fig:minimal_model}.
The linear spring represents the bending stiffness of the leaf spring, which connects the masses $m_1$ and $m_2$, whereas the nonlinear element represents the bolted joint connection between the masses $m_2$ and $m_3$.
We focus on the lowest-frequency elastic translational mode, of which the shape is schematically indicated by the blue arrows in Fig.\ \ref{fig:minimal_model}.

\begin{figure}
\centering
\includegraphics[width=0.49\textwidth]{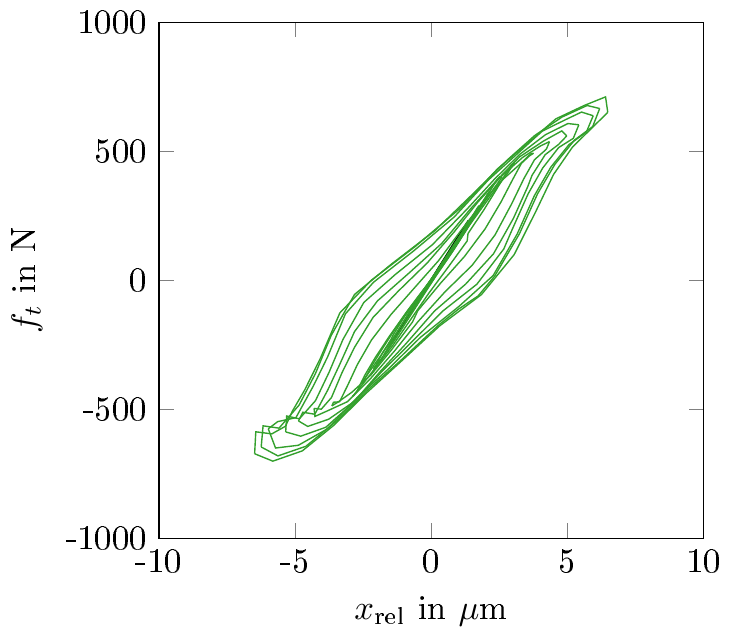}
	\caption{Frictional hysteresis cycles identified from the joint resonator.}
	\label{fig:hyst_damp_loop}
\end{figure}

An important feature of the joint resonator is that the axial force $f_{\mathrm t}$ of the bolted joint can be estimated by the balance of linear momentum of the oscillator $m_3$ when its mass and acceleration are measured; \ie $f_{\mathrm t} = m_3\ddot x_3$.
The relative deformation $x_{\mathrm{rel}}$ of the bolted joint is determined by integrating in the frequency domain the relative acceleration $\ddot x_{\mathrm{rel}} = \ddot x_3-\ddot x_2$ of the oscillators adjacent to the joint.
The axial force and relative displacement form the hysteresis loops depicted in \fref{hyst_damp_loop}.
For low relative displacement amplitudes, the bolted joint acts mainly as linear spring, and the hysteresis loops degenerate to straight lines.
For higher relative displacement amplitudes, the friction joint is first driven into a micro-slip regime, and the hysteresis loop encloses an area corresponding to the energy dissipated by friction.
The effective slope of the hysteresis varies with the excitation level, indicating a change in effective stiffness.
Thus, the bolted friction joint causes both damping and stiffness nonlinearity.

Besides accelerometers, the force transduced to the system via the shaker is measured with a load cell.
The dSPACE DS1103 rapid prototyping system was used for the implementation of the force controller with the parameters specified in \ref{append:pll}.
All sensor signals could in principle be recorded with this system, too. To reduce the computational effort for the dSPACE hardware, all signals were instead recorded using a DEWEBook USB2-8, \cf \fref{exp_setup}. Thus, a sampling rate of both controller and data acquisition of 10kHz was achieved.

\subsection{Modal characteristics}
\begin{figure}[tb]
	\centering
	\begin{subfigure}[]{0.49\textwidth}
		\centering
		\includegraphics{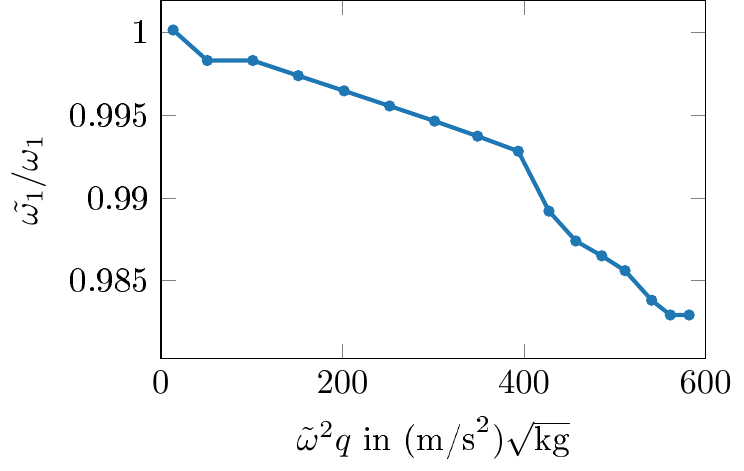}\caption{}\label{fig:FAP_DAP_freq}
	\end{subfigure}
	\begin{subfigure}[]{0.49\textwidth}
		\centering
		\includegraphics{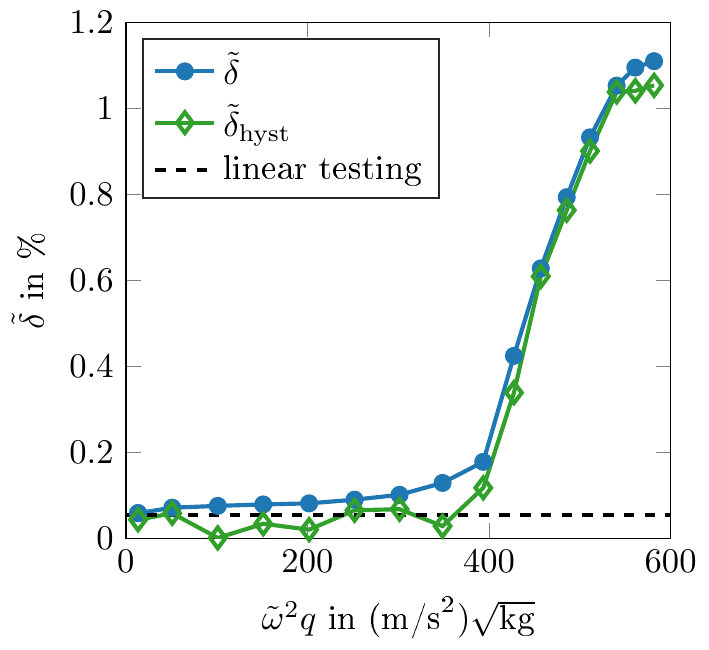}\caption{}\label{fig:FAP_DAP_damp}
	\end{subfigure}
\caption{Nonlinear modal properties vs. vibration level: (a) natural frequency (b) modal damping ratio.}
\end{figure}

The modal testing of the underlying linear modes was carried out with LMS Scadas mobile using random shaker excitation from 10 Hz to 2000 Hz.
The nonlinear modal testing involved 16 successively increasing excitation levels.
For each increment, the excitation level is held constant for 25s. This is sufficient for the PLL controller to reach a locked state and to record the last 80 steady-state excitation periods for modal property extraction.
The results are depicted in Fig.\ \ref{fig:FAP_DAP_freq}. The modal frequency decreases with increasing vibration level, reflecting the well-known softening effect of friction joints. It is noted that the decrease already occurs in the low energy regime where such stiffening nonlinearities are unexpected.
The experiments were repeated several times and the decreasing modal frequency for low energies was observed in all measurements. However, only the results of the last measurement are shown here.
At around $\tilde{\omega}^2 q=400 (\text{m}/\text{s}^2)\sqrt{\text{kg}}$, a comparatively sharp bend can be seen both in the modal frequency and the damping characteristic, which is typical for the transition from mostly sticking to micro-slip behavior \cite{gaul2001,bogr2011}.

A known alternative to the proposed method is to quantify an equivalent damping ratio of a periodic resonant vibration as \cite{Ungar1962,bogr2011}
\begin{equation}
\tilde{\delta}_{\rm hyst} = \frac{W^{\rm hyst}_{\rm diss}}{4\pi E^{\rm max}_{\rm pot}}.
\label{eq:loss_factor}
\end{equation}
Herein, $W^{\rm hyst}_{\rm diss}$ is the dissipated work per cycle by friction, which equals the area, $W^{\rm hyst}_{\rm diss} = \oint f_{\rm t} \text{~d} x_{\rm rel}$, enclosed in the hysteresis loop (\cf \fref{hyst_damp_loop}).
The maximum potential energy $E^{\rm max}_{\rm pot}$ can be determined as
\begin{equation}
E^{\rm max}_{\rm pot} = \int_{0}^{x_{\rm rel,max}}f_{\rm t} \dd x_{\rm rel} + \int_{0}^{x_{\rm k,max}} f_{\mathrm k} \dd x_{\rm k},
\label{eq:Epot_max}
\end{equation}
where $x_{\rm k}=x_2-x_1$ and the force $f_{\rm k}$ in the leaf spring is determined by the balance of linear momentum of oscillator $m_1$, $f_{\rm k} = m_1 \ddot{x}_1-f_{\rm exc}$, using the measured mass $m_1$, acceleration $\ddot x_1$ and excitation force $f_{\mathrm{exc}}$.
\eref{Epot_max} holds under the assumption that the oscillators move in unison, so that joint and leaf spring reach their maximum potential energies simultaneously, and $E_{\rm pot}=0$ for $x_{\rm k}=0=x_{\mathrm{rel}}$.
As can be seen in \fref{FAP_DAP_damp}, hysteresis based ($\tilde{\delta}_{\rm hyst}$) and modal damping ratio ($\tilde\delta$) have a qualitatively similar dependence.
$\tilde\delta_{\rm hyst}$ is slightly lower than $\tilde{\delta}$, which indicates that the identified frictional hysteresis in \fref{hyst_damp_loop} does not capture all dissipation mechanisms within the system. Additional dissipation sources are frictional dissipation attributed to non-axial loading direction of the bolted joint, frictional dissipation due to other joints, and material as well as aerodynamic damping.
It should be emphasized that the hysteresis-based damping ratio $\tilde{\delta}_{\rm hyst}$ can only be evaluated in a straight-forward way for lumped parameter systems such as the joint resonator.
In contrast, the modal testing procedure can be readily applied to distributed parameter systems.

\begin{figure}
	\begin{subfigure}[]{0.49\textwidth}
		\centering
		\includegraphics{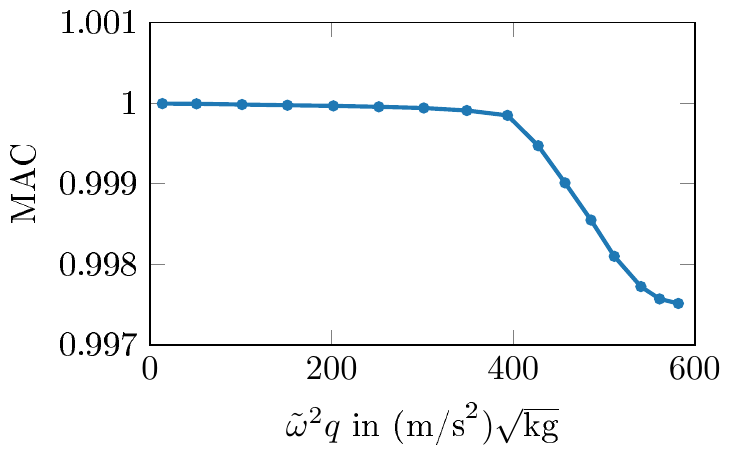}\caption{}\label{fig:MAC}
	\end{subfigure}
	\begin{subfigure}[]{0.49\textwidth}
		\centering
		\includegraphics{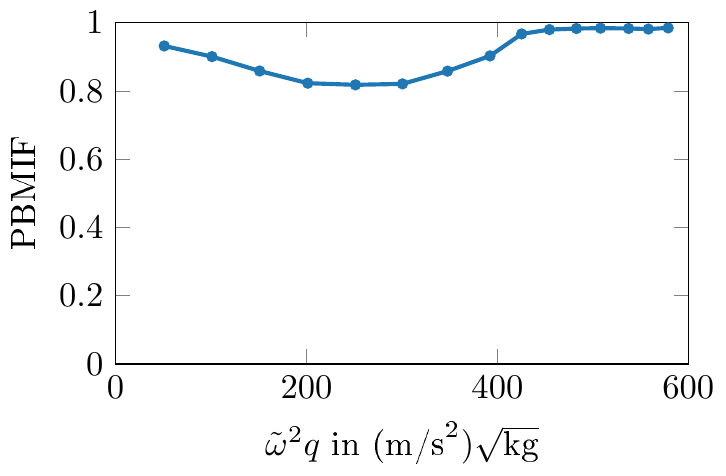}\caption{}\label{fig:PBMIF}
	\end{subfigure}
\caption{(a) correlation (MAC) between fundamental harmonic of nonlinear mode shape and associated linear mode shape, (b) \PBMIF. }
\end{figure}

The modal assurance criterion (MAC),
\begin{equation}
MAC = \frac{\left\|\bs{\phi}^{\rm T} \tilde{\bs{\phi}}_1 \right\|^2}{\bs{\phi}^{\rm T}\bs{\phi}\tilde{\bs{\phi}}^{\rm H}_1 \tilde{\bs{\phi}}_1 },
\label{eq:MAC}
\end{equation}
measures the correlation between the fundamental harmonic component $\tilde{\bs{\phi}}_1$ of the nonlinear mode shape and the shape $\bs\phi$ of the underlying linear mode. It is depicted in Fig.\ \ref{fig:MAC}.
The MAC remains very close to unity throughout the tested range of vibration levels, indicating only minor variation of the vibrational deflection shape. Still, the high consistency with the results of the linear modal analysis should be emphasized. Moreover, a distinct qualitative dependence of the MAC on the vibration level can be ascertained, which is consistent with the evolution of the modal frequency and damping properties.

The \PBMIF is depicted in Fig.\ \ref{fig:PBMIF}. Interestingly, it assumes highest values in the micro-slip regime, while it takes on values lower than those in the numerical study in the intermediate linear regime. Since no reference is available in this case, it remains unclear how well the \PBMIF correlates with the quality of the extracted modal properties.

\subsection{Indirect verification with frequency response measurements}
As no high fidelity reference is available for the extracted modal characteristics, they have to be further verified indirectly.
To this end, their relevance for representing the near-resonant frequency response is assessed.
This is achieved by comparing measured frequency responses to numerically synthesized ones based only on the extracted modal characteristics.
For the synthesis, it is assumed that the system's response is dominated by the extracted nonlinear mode, which leads to the equation \cite{szem1979}
\begin{equation}
[- \Omega^2+2 \ii \Omega \tilde{\omega}\tilde{\delta} +\tilde{\omega}^2] q\ee^{\ii\Delta\Theta} =    \tilde{\bs{\phi}}_{1}^{\rm H} \bs{f}_{1,\rm exc}.\label{eq:nmsdof}
\end{equation}
Herein, $\Omega$ is the frequency and $\bs{f}_{1,\rm exc}$ is the fundamental harmonic of the external forcing, and $q$ and $\Delta\Theta$ are the magnitude and the phase of the modal coordinate.
Note that \eref{nmsdof} corresponds to the forced response equation of a single nonlinear modal oscillator.
In addition to the response of the nonlinear mode, the contribution of the remaining linear modes is linearly superimposed using the low and high frequency residuals identified from the linear modal testing step.
For details on the frequency response synthesis, the reader is referred to \cite{Peter.2018, krac2013a}.
As reference, slow sine sweep responses are measured, with a sweep rate of 0.2 Hz/s and controlled excitation level, using the controller Br\"uel \& Kj\ae r Vibration Exciter Control Type 1050.
The frequency range of 350 and 390 Hz is tested for two different excitation levels $f_{1,\rm exc}=38.15~\mathrm{N}$ and $14.65~\mathrm{N}$.

\begin{figure}
	\center
	\includegraphics{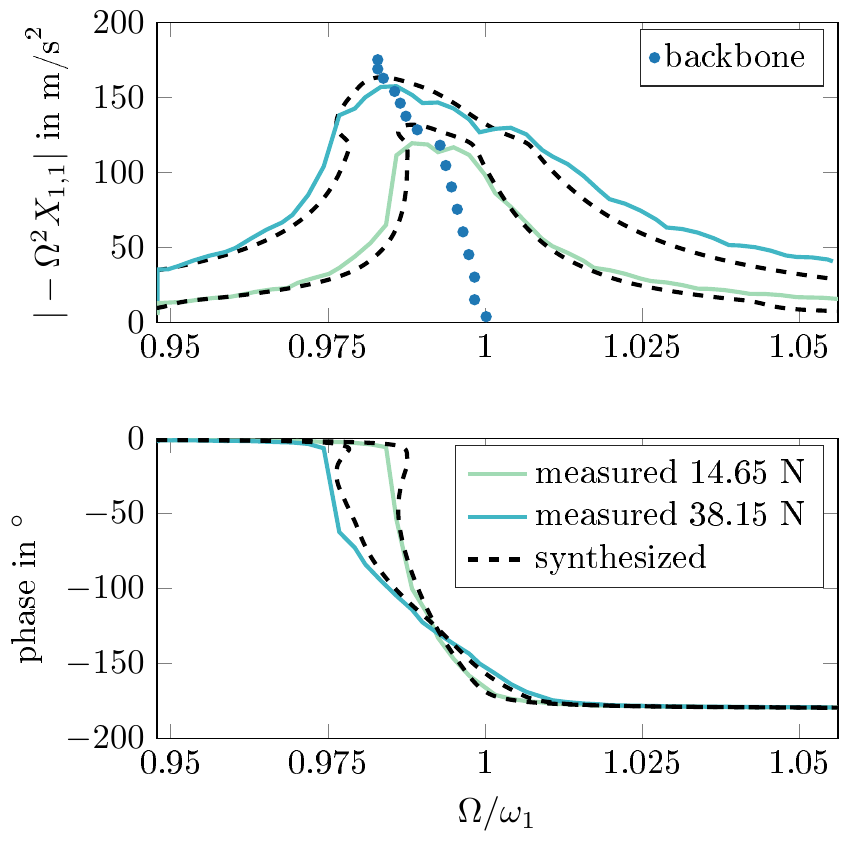}
	\caption{Measured and synthesized frequency response of acceleration $\ddot x_1$ for two different excitation levels: (a) amplitude response, (b) phase response.}
	\label{fig:FRF_synth}
\end{figure}

The results of the synthesis and the reference measurement are shown in Fig.\ \ref{fig:FRF_synth}. It can be seen that synthesized and measured response agree well in the vicinity of the resonance, indicating high accuracy of the extracted modal properties. Some deviation can be seen around the peak which may be attributed to some small error in the modal damping estimation or imperfections controlling the force level of the sine sweep reference in this frequency range which have been observed in the measurements.

Based on the experience gathered throughout testing the joint resonator, it was found that the proposed nonlinear modal testing method is comparatively fast and robust against the specific settings of the excitation controller. In contrast, the frequency response measurements were found quite sensitive to the sweep rate. Low sweep rates were needed to achieve robust results and a fairly constant force level over the entire frequency range. This resulted in comparatively high measurement times of 200 s per excitation level. For the modal testing, the chosen number of points lead to a measurement duration of 390 s, being less time consuming if forced responses at several excitation levels are of interest, \eg parameter studies, which agrees with the observations presented in \cite{Peter.2018}.

\section{Conclusions\label{sec:conclusions}}
The developed method can be viewed as a novel experimental procedure for nonlinear modal testing which has the unique feature that it can also be applied to damped structures.
The method permits the extraction of the modal frequency, damping ratio and deflection shape (including harmonics), for each mode of interest, as function of the vibration level.
A particularly simple experimental realization was investigated: a single-point forcing whose fundamental frequency component is in phase resonance with the vibration response at this point.
Thus, the method effectively tracks the backbone of nonlinear frequency responses.
It therefore requires only a single response per vibration level, while it still captures the underlying near-resonant dynamics. This makes the method more time-efficient and less susceptible to destructive testing under long exposure to high vibrations, as compared to alternative vibration testing methods.
The accuracy of the method was thoroughly validated for a virtual experiment against the direct computational nonlinear modal analysis.
Moreover, the experimental application to a friction-damped system demonstrated its readiness and usefulness for real-world applications.
In accordance with the extended periodic motion concept, the modal properties are extracted from steady-state time series. This leads to high robustness against measurement noise, as compared to damping quantification methods that analyze the free decay.
The method requires only minimal prior knowledge of the system to be tested, which is a substantial advantage over most experimental methods, including hysteresis-based damping quantification which is only applicable to lumped parameter systems.
Another advantage of the method is that it can be implemented using standard equipment (shaker, accelerometers, PLL controller).

In future studies, it would be interesting to apply the proposed method to real-life engineering systems, to further explore its performance as compared to more conventional methods involving stepped or swept sines.
During the preparation of this article, we already further verified the method experimentally for a jointed beam \cite{Scheel2018} and numerically validated it for a model of shrouded turbine blades \cite{Scheel2017}.
Moreover, the range of utility of the method should be explored for more complicated systems, including multiple or stronger nonlinear elements and more closely-spaced modes.
It is believed that the single-point, single-frequency excitation control will have to be generalized to multi-point, multi-frequency excitation control in order to test some of these systems.
Furthermore, the advantages and limitation of the proposed method will be evaluated through comparison with other nonlinear system identification approaches, such as NSID.
Another interesting perspective is the application of the method to self-excited systems.

\appendix
\setcounter{figure}{0}
\setcounter{table}{0}

\section{Active and apparent power of a self-excited system} \label{append:power_selfexc}

If for simplicity the system is transferred to (linear) mass normalized modal coordinates $\bs x(t) = \bs{\Phi} \bs\eta(t)$, the instantaneous power provided by the negative damping term $-\xi\mbf M\dot{\bs x}$ can be written as a sum over all $N_l$ modes,
\begin{equation}
p(t) = \left( -\xi\mbf M\dot{\bs x}(t)\right)^{\rm T} \dot{\bs x}(t) = -\xi \dot{\bs \eta}(t)^{\rm T} \bs{\Phi}^{\rm T} \mbf M \bs{\Phi} \dot{\bs \eta}(t) = -\xi \dot{\bs \eta}(t)^{\rm T} \dot{\bs{\eta}}(t) = -\xi  \left( \sum_{l=1}^{N_l} \dot{\eta}_l(t) \dot{\eta}_l(t)\right). \label{eq:inst_power_self_2}
\end{equation}
The active power for a mode $l$ reads (\cf \eref{activePower})
\begin{equation}
P^{\rm se}_l  = \real{-\sum\limits_{n=1}^{\infty} \frac12 \xi V_{l,n} \overline{V}_{l,n}} = -\sum_{n=1}^\infty{\frac12 \xi \abs{V_{l,n}} \abs{V_{l,n}}} =  - \frac12 \xi \sum_{n=1}^\infty{V_{l,n}^2}.
\label{eq:act_power_fourier_se}
\end{equation}
Herein, $V_{l,n}$ denotes the $n$-th harmonic of $\dot{\eta}_l$. Furthermore, the negative damping force is proportional to the velocity, such that the apparent power of the self-excited system can be calculated as
\begin{equation}
S^{\rm se}_l = \sqrt{ \sum_{n=1}^\infty{\frac12 \xi^2 V_{l,n}^2}} \sqrt{\sum_{n=1}^\infty{\frac12 V_{l,n}^2}}= \frac12 \xi \sum_{n=1}^\infty{V_{l,n}^2} = -P^{\rm se}_l.
\end{equation}

\section{Phase-locked loop}\label{append:pll}

\begin{figure}
	\centering
	\includegraphics{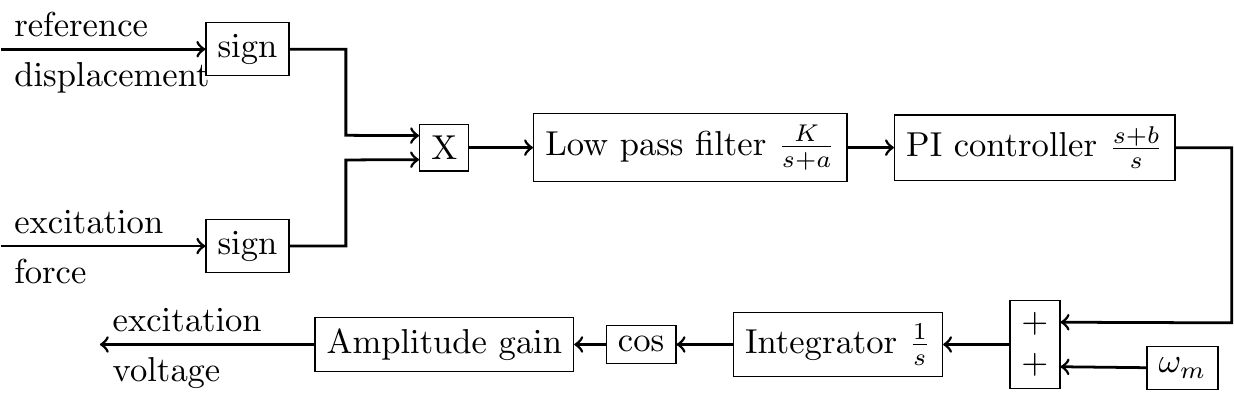}	
	\caption{Scheme of PLL controller with transfer functions.} \label{fig:PLL_scheme}
\end{figure}

In this work, the PLL as sketched in Fig.\ \ref{fig:PLL_scheme} is used. The corresponding parameters are stated in Tab.\ \ref{tab:PLL parameters num} and \ref{tab:PLL parameters exp}. For more details, the interested reader is referred to \cite{Peter.2017,Abramovitch2002}.

\begin{table}
	\begin{minipage}[b]{0.5\textwidth}
		\centering
		\begin{tabular}{l|c}
			\hline
			Parameter & Value\\
			\hline
			$K$ for $k_t = 8000 \text{ N/m}$ & $2 \pi$\\
			$K$ for $k_t > 8000 \text{ N/m}$ & $30 \pi$\\
			$a$ & $2 \pi$\\
			$b$ & $\pi$\\
			$\omega_m$ & first linear eigenfrequency\\
			\hline
		\end{tabular}
		\captionof{table}{Parameters of PLL used in the numerical study.}\label{tab:PLL parameters num}
	\end{minipage}
	\hfill	
	\begin{minipage}[b]{0.5\textwidth}
		\centering
		\begin{tabular}{l|c}
			\hline
			Parameter & Value\\
			\hline
			$K$ & $20 \pi$\\
			$a$ & $2 \pi$\\
			$b$ & $\pi$\\
			$\omega_m$ & $370$ Hz\\
			\hline
		\end{tabular}
		\captionof{table}{Parameters of PLL used for experiments.}\label{tab:PLL parameters exp}
	\end{minipage}
\end{table}

\section*{References}

\bibliography{jsv_scheel_ref}

\end{document}